# Mechanical properties of superhard boron subnitride $B_{13}N_2$

Vladimir L. Solozhenko [a, *]  and  Volodymyr Bushlya [b]

[a] LSPM–CNRS, Université Paris Nord, 93430 Villetaneuse, France

[b] Division of Production and Materials Engineering, Lund University, 22100 Lund, Sweden

Microstructure and mechanical properties of bulk polycristalline rhombohedral boron subnitride $B_{13}N_2$ synthesized by crystallization from the B–BN melt at 7 GPa have been systematically studied by micro- and nanoindentation, atomic force microscopy and scanning electron microscopy. The obtained data on hardness, elastic properties and fracture toughness clearly indicate that $B_{13}N_2$ belongs to a family of superhard phases and can be considered as a promising superabrasive or binder for cubic boron nitride.

Rhombohedral boron subnitride $B_{13}N_2$ has been recently synthesized by crystallization from the B-BN melt at 5 GPa [1-3]. The structure of $B_{13}N_2$ belongs to the *R*-3*m* space group and represents a new structural type produced by the distorted $B_{12}$ icosahedra linked together by N–B–N chains and inter-icosahedral B–B bonds [1]. Boron subnitride is refractory ($T_m$ = 2430(20) K at ambient pressure [4]) and low-compressible ($B_0$ = 200(15) GPa [5]) phase similar to other boron-rich solids with structures related to α-rhombohedral boron ($B_6O$, $B_4C$, etc.). According to the predictions made in the framework of thermodynamic model of hardness [6,7], $B_{13}N_2$ is expected to exhibit hardness of 40 GPa [8] comparable to that of commercial polycrystalline cubic boron nitride. Here we present the results of the comprehensive study of mechanical properties of $B_{13}N_2$ boron subnitride.

Well-sintered bulks of polycrystalline $B_{13}N_2$ has been synthesized in a toroid-type apparatus with a specially designed high-temperature cell [9] at 7 GPa by quenching of the B–BN melt from 2630 K in accordance with high-pressure phase diagram of the B–BN system [10]. Powders of crystalline β-rhombohedral boron (99%, Alfa Aesar) and hexagonal graphite-like boron nitride (hBN) (99.8%, Johnson Matthey GmbH) have been used as starting materials. The X-ray diffraction study (TEXT 3000 Inel, CuKα1 radiation) has shown that the recovered bulk samples contain well-crystallized $B_{13}N_2$ ($a$ = 5.4585(8) Å, $c$ = 12.253(2) Å), in mixture with cubic BN (10-15 vol.%) due to the peritectic nature of the L + BN $\rightleftharpoons$ $B_{13}N_2$ reaction [4,10].

The recovered samples (cylinders 4-mm diameter and 3-mm height) were hot mounted in electrically conductive carbon-fiber reinforced resin, and were planar ground with 1200 grit SiC and subsequently polished with 9-μm and 1-μm diamond suspensions, followed by super-finishing with 0.04-μm $SiO_2$ colloidal solution. Extensive duration of the super-finishing (~1.5 hour) and low process pressure (0.02 MPa) ensured the minimal mechanical damage to the material surface after final polishing.

Microstructure of the polished samples have been studied by high-resolution scanning electron microscopy (SEM) using LEO/Zeiss 1560 microscope in secondary electron and InLens modes. SEM results indicate the presence of residual cubic BN (bright contrast) localized as individual inclusions (Fig. 1a). The dimensions of the $B_{13}N_2$ phase pools vary from 50 to 200 μm. Close-up image of the boron subnitride region taken with an InLens detector shows that $B_{13}N_2$ is polycrystalline with the grain size of 50 to 600 nm (Fig. 1b).

Microindentation has been performed using the Ernst Leitz Wetzlar microhardness tester under loads ranging from 0.25 to 6.0 N; at least, five indentations have been made at each load. The indent sizes were measured with a Leica DMRME optical microscope under 1000× magnification in the phase contrast regime. Vickers hardness ($H_V$) was determined from the residual imprint upon indentation and was calculated following the standard definitions according to Eq. 1:

$$H_V = \frac{1.8544 \cdot P}{d^2} \tag{1}$$

where $P$ and $d$ are the applied load and residual imprint diagonal, respectively. The value of Knoop hardness ($H_K$) was determined by Eq. 2:

$$H_K = \frac{P}{0.070279 \cdot d^2} \tag{2}$$

where P is the applied load and $d$ is the length of a large diagonal of an indent.

Vickers hardness measurements have shown that the calculated microhardness abruptly decreases with the load and reaches the asymptotic value of $H_V = 41(2)$ GPa already at a load of 1 N (Fig. 2a). The experimental Vickers hardness of $B_{13}N_2$ is in an excellent agreement with the value previously calculated in the framework of the thermodynamic model of hardness [8], and is comparable to the hardness of commercial polycrystalline cubic boron nitride. The load dependence of calculated Knoop hardness is presented in Fig. 2b; the asymptotic hardness value is $H_K = 32(1)$ GPa.

Nanoindentation study has been performed on Micro Materials NanoTest Vantage system with trigonal Berkovich diamond indenter (the tip radius of 120 nm). The maximal applied load was 500 mN. Loading at the rate of 0.5 mN/s was followed by a 10 s holding and unloading at the same rate. AFM microscope Dimension 3100 (Digital Instruments) in tapping mode was used on nanoindentation imprints for pile-up correction. Fig. 3 shows a characteristic load-displacement curve for bulk $B_{13}N_2$.

Evaluation of the hardness and elastic modulus was performed in accordance to the Oliver-Pharr method [11]. The hardness of the sample was determined by Eq. 3:

$$H_N = \frac{P_{max}}{A(h_c)} \tag{3}$$

where $P_{max}$ is the maximum applied load and $A(h_c)$ is the projected contact area. The area function $A(h_c)$ was calibrated on a standard fused silica reference sample. Correction of the area function for the pile-up effects was based on the indent topography data obtained on the actual samples by atomic force microscopy.

From 7 independent nanoindentation experiments the nanohardness of $B_{13}N_2$ was found to be $H_N = 36(2)$ GPa that is in a good agreement with our microhardness data. The elastic recovery of $B_{13}N_2$ was determined as the ratio of elastic work to the total work of the indentation by Eq. 4:

$$R_W = \frac{W_e}{W_p + W_e} \times 100\%$$ (4)

where $W_p$ and $W_e$ are plastic and elastic works, respectively. From experimentally found values $W_p$=77(9) nJ and $W_e$=97(1) nJ the elastic recovery $R_W$ has been estimated as 55(6)% which is slightly lower than that of single-crystal cubic BN (60% [12]).

Reduced modulus $E_r$ was determined from stiffness measurements that are governed by elastic properties of the sample and diamond indenter via Eq. 5:

$$E_r = \left( \frac{1 - \nu_s^2}{E_s} + \frac{1 - \nu_i^2}{E_i} \right)^{-1}$$ (5)

where $E_s$, $E_i$ are Young's moduli and $\nu_s$, $\nu_i$ are the Poisson's ratios of the sample and indenter, respectively. The elastic modulus of the material hence can be calculated for known properties of diamond ($E_i = 1141$ GPa and $\nu_i = 0.07$ [11]) and Poisson's ratio of the sample. The Young's modulus of $E = 515(16)$ GPa was calculated by Eq. 5 using the experimental $E_r = 365(8)$ GPa value (data from 7 independent nanoindentation experiments) under the assumption that Poisson's ratio of $B_{13}N_2$ is equal to $\nu = 0.23$ (theoretically predicted using the Voigt-Reuss-Hill approach [13]). The variation of $\nu$ over the 0.16-0.28 range results in variation of $E$-value from 530 to 501 GPa which is within the experimental error of Young's modulus evaluation from the present set of nanoindentation data. This allows the conclusion that the theoretically predicted Young's modulus value (387 GPa [13]) is strongly underestimated.

Using the relation (6) between Young's ($E$) and shear ($G$) moduli for an isotropic material

$$G = \frac{E}{2(1 + \nu)}$$ (6)

and Poisson's ratio $\nu = 0.23$ [13], the shear modulus of $B_{13}N_2$ was evaluated as $G = 209(6)$ GPa which is significantly higher than the theoretically predicted values (157 GPa [13] and 162 GPa [14].

The fracture toughness ($K_{Ic}$) was studied with an Ernst Leitz Wetzlar microhardness tester using a Vickers diamond indenter under 6 N load. The lengths of radial cracks emanating from the indent corners were measured in polarized light with Alicona InfiniteFocus 3D optical microscope under 1000× magnification. The value of $K_{Ic}$, was determined in terms of the indentation load $P$ and the mean length (surface tip-to-tip length $2c$) of the radial cracks according to Eq. 7 [15]:

$$K_{Ic} = x_v \cdot (E/H_V)^{0.5} \ (P/c^{1.5})$$ (7)

where $x_v = 0.016(4)$, $E$ is Young's modulus and $H_V$ is load-independent Vickers hardness. The average fracture toughness of $B_{13}N_2$ has been estimated as $K_{Ic} = 1.9(4)$ MPa·m$^{\frac{1}{2}}$ which is 33% lower than the 2.8 MPa·m$^{\frac{1}{2}}$ value for single-crystal cBN [16]. As the crack lengths (25–40 μm) are much longer than the grains (Fig. 4), the $K_{Ic}$ value is a characteristic of the bulk $B_{13}N_2$ material as a whole.

The data on mechanical and elastic properties of $B_{13}N_2$ are summarized in the Table. Due to high hardness and elastic recovery as well as high thermal stability, oxidation resistance and adhesion to boron nitride, rhombohedral boron subnitride $B_{13}N_2$ offers promise as a potential superabrasive or binder for cubic boron nitride.

The authors thank Dr. Vladimir A. Mukhanov for assistance in high-pressure synthesis of $B_{13}N_2$, Prof. Jinming Zhou for nanoindentation study and Prof. Jan-Eric Ståhl for helpful discussions. This work was financially supported by the European Union's Horizon 2020 Research and Innovation Programme under the Flintstone2020 project (grant agreement No 689279).

Table  Hardness, elastic moduli and fracture toughness of superhard boron nitrides

| | $H_V$ | $H_K$ | $H_N$ | $E$ | $G$ | $B$ | $K_{Ic}$ |
|---|---|---|---|---|---|---|---|
| | GPa | | | | | | MPa·m$^{1/2}$ |
| $B_{13}N_2$ | 41(2) | 32(1) | 37(1) | 515(16) | 209(6) | 200 [5] | 1.9(4) |
| cubic BN | 62 [a] | 44 [a] | 55 [a] | 909 [b] | 409 [17] | 397 [18] | 2.8 [16] |

[a]  single crystal, (111) face [12]

[b]  polycrystalline material [12]

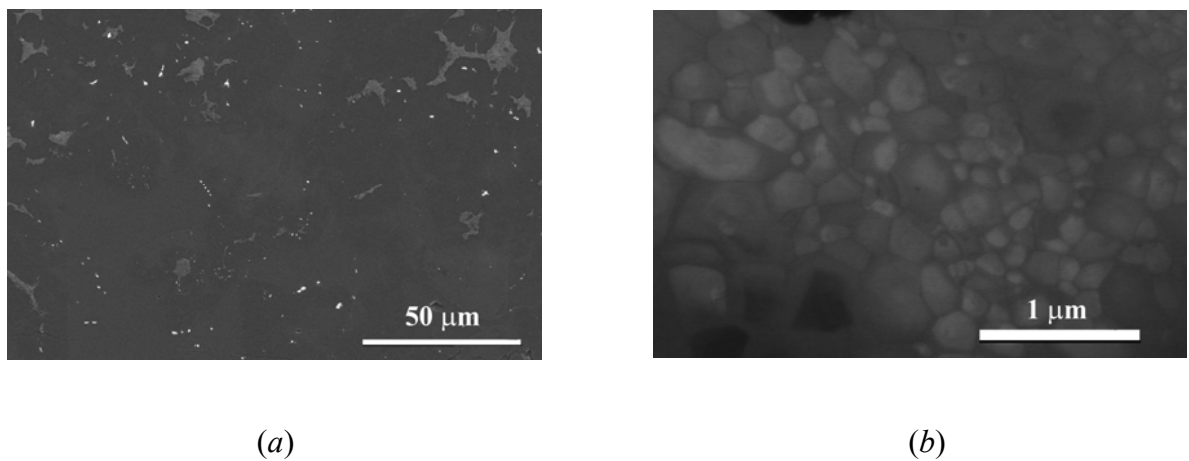

<div align="center">(<em>a</em>)        (<em>b</em>)</div>

Fig. 1  SEM overview of the sample microstructure (<em>a</em>) and an InLens close-up
image of the region of pure $B_{13}N_2$ (<em>b</em>).

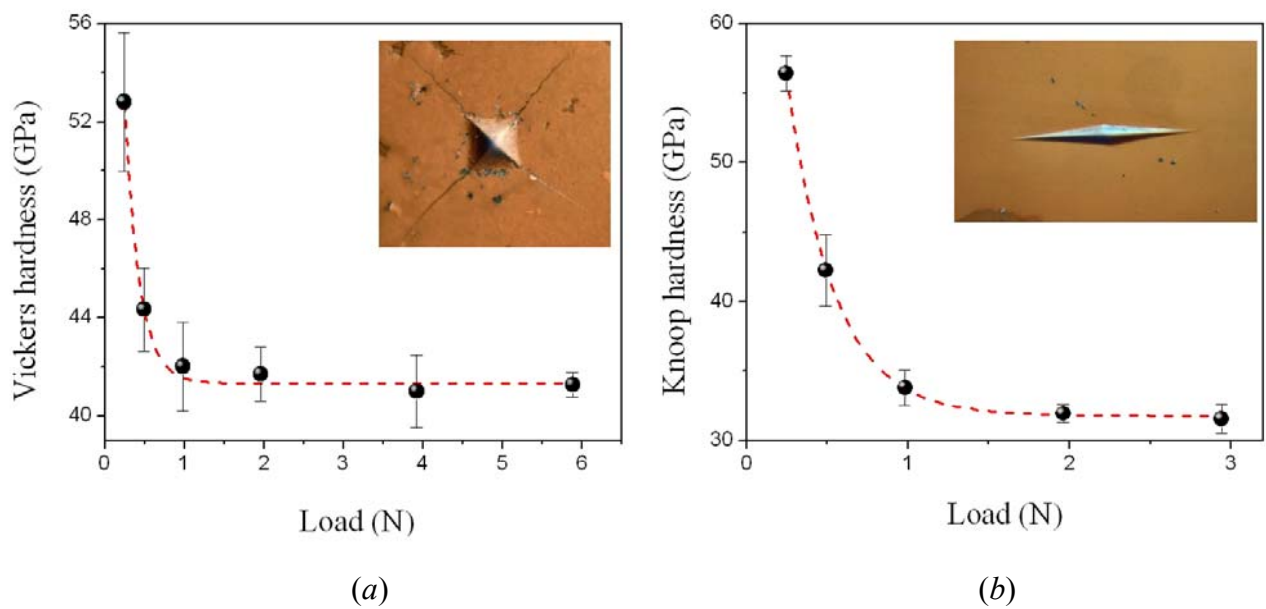

Fig. 2     Vickers (*a*) and Knoop (*b*) microhardness of bulk boron subnitride $B_{13}N_2$ *vs* load.
Insets: optical microscope images of the indents formed by Vickers and Knoop
indenters under loads of 6 N and 3 N, respectively.

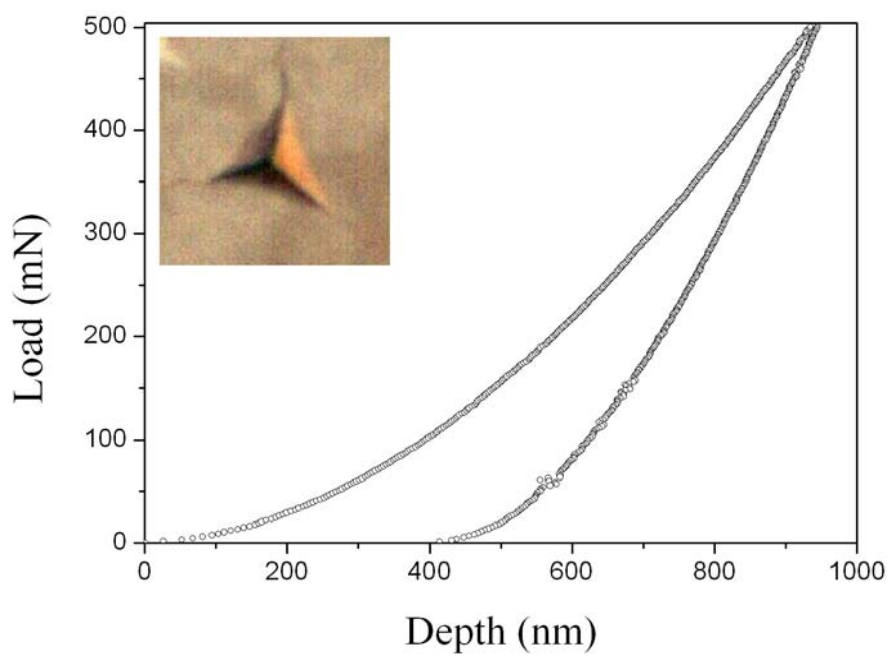

Fig. 3    Characteristic load-displacement curve for bulk boron subnitride $B_{13}N_2$.
Inset: optical microscope image of the indent formed by Berkovich
indenter under load of 500 mN.

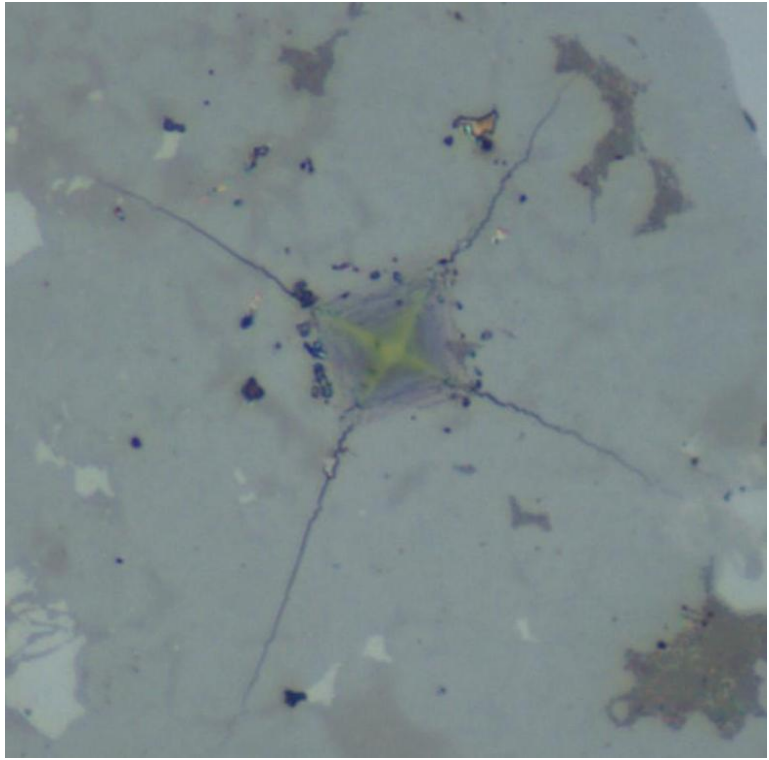

Fig. 4    Optical microscope image of the radial crack system in polycrystalline $B_{13}N_2$ under indentation fracture toughness test at 6-N load;  width of field is 65 μm.